\documentclass{jfm}
\usepackage{mathrsfs,eucal}
\usepackage{graphicx}
\usepackage[T1]{fontenc}

\let\realverbatim=\verbatim
\let\realendverbatim=\endverbatim
\renewcommand\verbatim{\par\addvspace{6pt plus 2pt minus 1pt}\realverbatim}
\renewcommand\endverbatim{\realendverbatim\addvspace{6pt plus 2pt minus 1pt}}
\makeatletter
\newcommand\verbsize{\@setfontsize\verbsize{10}\@xiipt}
\renewcommand\verbatim@font{\verbsize\normalfont\ttfamily}
\makeatother

\ifCUPmtlplainloaded \else
  \checkfont{eurm10}
  \iffontfound
    \IfFileExists{upmath.sty}
      {\typeout{^^JFound AMS Euler Roman fonts on the system,
                   using the 'upmath' package.^^J}%
       \usepackage{upmath}}
      {\typeout{^^JFound AMS Euler Roman fonts on the system, but you
                   don't seem to have the}%
       \typeout{'upmath' package installed. JFM.cls can take advantage
                 of these fonts,^^Jif you use 'upmath' package.^^J}%
       \providecommand\upartial{\upartial}%
      }
  \else
    \providecommand\upartial{\upartial}%
  \fi
\fi

\ifCUPmtlplainloaded \else
  \checkfont{msam10}
  \iffontfound
    \IfFileExists{amssymb.sty}
      {\typeout{^^JFound AMS Symbol fonts on the system, using the
                'amssymb' package.^^J}%
       \usepackage{amssymb}%
       \let\le=\leqslant  
       \let\ge=\geqslant  
      }{}
  \fi
\fi

\ifCUPmtlplainloaded \else
  \IfFileExists{amsbsy.sty}
    {\typeout{^^JFound the 'amsbsy' package on the system, using it.^^J}%
     \usepackage{amsbsy}}
    {}
\fi

\newcommand{\df}[2]{\displaystyle\frac{#1}{#2}}
\newcommand{\tf}[2]{\textstyle\frac{#1}{#2}}

\newcommand{\be}{\begin{eqnarray}}
\newcommand{\en}{\end{eqnarray}}
\newcommand{\no}{\nonumber}

\newsavebox{\astrutbox}
\sbox{\astrutbox}{\rule[-5pt]{0pt}{20pt}}

\title[Stability of Fully Nonlinear Stokes Waves]
{Stability of Fully Nonlinear Stokes Waves on Deep Water: Part 1. Perturbation Theory}

\author[S. G. Sajjadi \& D. L. Ross]%
{S\ls H\ls A\ls H\ls R\ls D\ls A\ls D\ns G.\ns
S\ls A\ls J\ls J\ls A\ls D\ls I \and
D\ls A\ls V\ls I\ls D\ns L.\ns
R\ls O\ls S\ls S}

\affiliation{Division of Geophysics and Planetary Physics,\\
CAMAC, Department of Applied Mathematics,\\
Embry-Riddle Aeronautical University,\\
Florida 32114, USA.}

\pubyear{2009}
\volume{538}
\pagerange{1--11}
\date{22 May 2009}
\begin{document}

\label{firstpage}
\maketitle

\begin{abstract}
{\footnotesize We consider a full set of harmonics for the Stokes
wave in deep water in the absence of viscosity, and examine the
role that higher harmonics play in modifying the classical
Benjamin-Feir instability.  Using a representation of the wave
coefficients due to Wilton, a perturbation analysis shows that the
Stokes wave may become unbounded due to interactions between the
$N^{th}$ harmonic of the primary wave train and a set of harmonics
of a disturbance.  If the frequency of the $n^{th}$ harmonic is
denoted  $ \omega _{n} =\omega \left( {1 \pm \delta } \right)$
then instability will occur if \be 0<\delta <\frac{\sqrt 2\
k\,n^ns_n }{\left( {n-1} \right)!} \en
 subject to the disturbance initially having sufficiently large
 amplitude.  We show that, subject to initial conditions, all
 lower harmonics will contribute to instability as well, and we
 identify the frequency of the disturbance corresponding to maximum growth rate.}
\end{abstract}

\section{Introduction}

In Benjamin-Feir's seminal work identifying criteria for the instability of
gravity waves on deep water [1], a bound was established on the frequency of
the disturbance as a function of wave steepness. This result was obtained
through a perturbation analysis, with the wave train represented by a second
order Stokes wave of the form
\[
\eta \left( {x,t} \right)=a\cos \left( {kx-\omega t}
\right)+\tf{1}{2}a^2k\cos \left[ {2(kx-\omega t)} \right]
\]
A disturbance was introduced in the form of two sidebands with slightly
different wave numbers and amplitude varying in time,
\be
 \tilde {\eta }_{1,2} \left( {x,t} \right)&=&\varepsilon _{1,2} \cos \left[
{k(1\pm \kappa )x-\omega \left( {1+\delta } \right)t+\gamma _{1,2} (t)}
\right]\no \\
        &+&\varepsilon _{1,2} ak\cos \left[ {kx+k(1\pm \kappa )x-\omega t-\omega
\left( {1+\delta } \right)t+\gamma _{1,2} \left( t \right)} \right] \no
\en
with the positive sign corresponding to the index $i=1$ and the negative
sign to $i=2$. The phase functions $\gamma _1 \left( t \right)$ and $\gamma
_2 \left( t \right)$ were necessary to insure that the dispersion relation
holds to second order for the sidebands.

Benjamin and Feir demonstrated that interactions between the disturbance and
the first harmonic of the primary wave can lead to resonance effects so long
as $0<\delta <\sqrt 2 \,ak$, (i.e., the frequency of the disturbance lies
sufficiently close to the fundamental frequency of the wave train). This
result followed by obtaining a set of three coupled ordinary differential
equations in $\varepsilon _1 \left( t \right)$, $\varepsilon _2 \left( t
\right)$, and $\theta \left( t \right)=\gamma _1 \left( t \right)+\gamma _2
\left( t \right)$ and demonstrating that the amplitudes $\varepsilon _1
\left( t \right)$ and $\varepsilon _2 \left( t \right)$ may become unbounded
if $\delta $ is in the stated range.

In this work, we consider the fully nonlinear Stokes wave
\[
\eta \left( {x,t} \right)=H=\sum\limits_{n=1}^\infty {a_n \cos \left( {k_n
x-\omega _n t} \right)=} \sum\limits_{n=1}^\infty {a_n \cos \zeta _n }
\]
and a disturbance represented by the infinite sum
\[
\tilde {\eta }_i =\sum\limits_{n=1}^\infty {\varepsilon _{n,i} \cos \zeta
_{n,i} +\sum\limits_{n=1}^\infty {\varepsilon _{n,i} a_n \cos \left( {\zeta
_n +\zeta _{n,i} } \right)\,\,\,} } .
\]
Out of all possible interactions between the disturbance modes and
the primary wavetrain, we show that critical mode interactions
with the primary wavetrain may be identified for \textit{every}
harmonic of the disturbance. Characterizing each of these
interactions leads to a coupled system of equations quite similar
to those considered by Benjamin and Feir. This leads to a range of
frequencies for the $n^{th}$harmonic of the sidebands where
contributions to instability may occur if $0<\delta
<\sqrt{2}kn^ns_n/(n-1 )!$, where $s_n $ is a parameter related to
wave steepness \textit{ak}. The value of $\delta $ corresponding
to marginal stability decreases with $n$, so that if instability
corresponding to a particular harmonic is identified, all lower
harmonics will contribute to instability as well, providing that
certain initial conditions are satisfied. The initial conditions
require $\left| {\varepsilon _{n,i} \left( {t_0 } \right)}
\right|$ to be sufficiently large at some fixed time $t_0 $.

\section{Formulation of the problem}

We consider a two-dimensional Stokes wave on water of finite depth, modeled
as an inviscid fluid, where $z=0$ represents the mean surface level, and
$z=\eta \left( {x,t} \right)$ the free surface. Then Laplace's equation
governs the motion, with nonlinearities captured through the kinematic and
dynamic conditions at the surface. We seek the velocity potential $\phi
\left( {x,z,t} \right)$ and $\eta \left( {x,t} \right)$ satisfying
\be
\begin{array}{l}
 \phi _{xx} +\phi _{zz} =0 \\

 \left. {\begin{array}{l}
 \phi _z =\eta _t +\varepsilon \phi _x \eta _x \,\,\,\,\,\, \\

 \phi _t +\eta +\frac{1}{2}\varepsilon \left( {\phi _x ^2+\phi _z ^2}
\right)=0 \\
 \end{array}} \right\} {\rm on}\,\,\,z=1+\varepsilon \eta \\

 \nabla \phi \to
0\,\,\,\,\,\,\,\,\,\,\,\,\,\,\,\,\,\,\,\,\,\,\,\,\,\,\,\,\,\,\,\,\,\,\,\,\,\,\,\,\,\,\,\,\,\,\,\,{\rm as}\,\,\,\,z\to
-\infty \\
 \end{array}\no
\en
We represent the free surface as
\[
\eta \left( {x,t} \right)=H=\sum\limits_{n=1}^\infty {a_n \cos \left( {k_n
x-\omega _n t} \right)=} \sum\limits_{n=1}^\infty {a_n \cos \zeta _n }
\]
And the potential as
\[
\phi \left( {x,y,t} \right)=\Phi =\sum\limits_{n=1}^\infty {\omega _n k_n
^{-1}a_n e^{k_n z}\sin \zeta _n }
\]
where $\omega_n ^2=gk_n $.

We introduce perturbations,
\begin{equation}
\label{eq1}
\eta =H+\varepsilon \tilde {\eta }
\quad
\phi =\Phi +\varepsilon \tilde {\phi }
\end{equation}
and seek conditions that allow the disturbance $\tilde {\eta }$ to exhibit
unbounded growth. By linearity of Laplace's equation, we have
\begin{equation}
\label{eq2}
\nabla ^2\tilde {\phi }=\tilde {\phi }_{xx} +\tilde {\phi }_{yy} =0
\end{equation}
$\nabla \tilde {\phi }\to 0$ as $z\to -\infty $.

The kinematic and dynamic boundary conditions may be written as
\begin{equation}
\label{eq3}
\eta _t +\eta _x \left( {\phi _x } \right)_{z=\eta } -\left( {\phi _z }
\right)_{z=\eta } =0
\end{equation}
\begin{equation}
\label{eq4}
g\eta +\left( {\phi _t } \right)_{z=\eta } +\tf{1}{2}\left( {\phi _x
^2+\phi _z ^2} \right)_{z=\eta } =0
\end{equation}
We conduct a standard perturbation analysis by substituting (\ref{eq1}) into (\ref{eq3}) and
(\ref{eq4}), setting $O\left( \varepsilon \right)$ terms to zero, and expanding the
resulting expressions about $z=H$. Then,
\begin{equation}
\label{eq5}
\tilde {\eta }_t +\tilde {\eta }_x \left( {\Phi _x } \right)_{z=H} +\tilde
{\eta }\left( {-\Phi _{zz} +H_x \Phi _{xz} } \right)_{z=H} +\left( {-\tilde
{\phi }_z +H_x \phi _x } \right)_{z=H} =0
\end{equation}
\begin{equation}
\label{eq6}
g\tilde {\eta }+\tilde {\eta }\left( {\Phi _x \Phi _{xz} +\Phi _z \Phi _{zz}
+\Phi _{tz} } \right)_{z=H} +\left( {\tilde {\phi }_t +\Phi _x \tilde {\phi
}_x +\Phi _z \tilde {\phi }_z } \right)_{z=H} =0
\end{equation}
Consider the analytic continuation of $\tilde {\phi }$ over the neighborhood
of the free surface, neglect terms greater than $O\left( {a_n ^2} \right)$,
and Taylor expand all expressions about the mean surface level $z=0$. Upon
substituting for $\Phi $ and $H,$ the kinematic condition (\ref{eq5}) may be written

\begin{equation}
\label{eq7}
\begin{array}{l}
 \tilde {\eta }_t +\tilde {\eta }_x \left( {\sum\limits_{n=1}^\infty {\omega
_n a_n \cos \zeta _n } +\sum\limits_{n=1}^\infty {\sum\limits_{m=1}^\infty
{a_n a_m \omega _n k_n \cos \zeta _n \cos \zeta _m } } } \right) \\
 \,\,\,\,\,\,\,+\tilde {\eta }\left( {-\sum\limits_{n=1}^\infty {\omega _n
k_n a_n \sin \zeta _n } -\sum\limits_{n=1}^\infty {\sum\limits_{m=1}^\infty
{a_n a_m k_n k_m \omega _m \sin \zeta _n \cos \zeta _m } } } \right) \\
 \,\,\,\,\,\,\,-\sum\limits_{n=1}^\infty {\sum\limits_{m=1}^\infty {a_n a_m
\omega _m k_m ^2\cos \zeta _n \sin \zeta _m}} \\
 \,\,\,\,\,\,\,-\sum\limits_{n=1}^\infty
{\sum\limits_{m=1}^\infty {\sum\limits_{q=1}^\infty {a_n a_m a_q k_m k_q
^2\omega _q \cos \zeta _n \sin \zeta _m \cos \zeta _q } } }   \\
 \,\,\,\,\,\,\,-\left( {\tilde {\phi }_z +\sum\limits_{n=1}^\infty {a_n k_n
\sin \zeta _n \tilde {\phi }_x } } \right)_{z=0} \\
 \,\,\,\,\,\,\,-\left(
{\sum\limits_{n=1}^\infty {a_n \cos \zeta _n } \tilde {\phi }_{zz}
+\sum\limits_{n=1}^\infty {\sum\limits_{m=1}^\infty {a_n a_m k_m \omega _m
\cos \zeta _n \sin \zeta _m } } \tilde {\phi }_{xz} } \right)_{z=0} \\
 \,\,\,\,\,\,\,-\frac{1}{2} \left(\sum\limits_{n=1}^\infty
{\sum\limits_{m=1}^\infty {a_n a_m \cos \zeta _n \cos \zeta _m } \tilde
{\phi }_{zzz}}\right)_{z=0}\\
 \,\,\,\,\,\,\,-\frac{1}{2} \left(\sum\limits_{n=1}^\infty {\sum\limits_{m=1}^\infty
{\sum\limits_{q=1}^\infty {a_n a_m a_q k_q \cos \zeta _n \cos \zeta _m \sin
\zeta _q } } } \tilde {\phi }_{xzz}  \right)_{z=0} =0 \\
 \end{array}
 \end{equation}

The dynamic condition (\ref{eq6}) may be written
\begin{equation}
\label{eq8}
\begin{array}{l}
 g\tilde {\eta }+\tilde {\eta }\left( \sum\limits_{n=1}^\infty
\sum\limits_{m=1}^\infty {a_n a_m \omega _n \omega _n k_m \cos \zeta _n
\cos \zeta _m } \right.\\
 \,\,\,\,\,\,\,\left.+\sum\limits_{n=1}^\infty {\sum\limits_{m=1}^\infty {a_n a_m
\omega _n \omega _n k_m \sin \zeta _n \sin \zeta _m } -}
\sum\limits_{n=1}^\infty {\omega _n ^2a_n \cos \zeta _n }  \right) \\
 \,\,\,\,\,\,\,+\sum\limits_{n=1}^\infty {\sum\limits_{m=1}^\infty
{\sum\limits_{q=1}^\infty {a_n a_m a_q k_q \omega _m \omega _q k_m k_q \cos
\zeta _n \cos \zeta _m \cos \zeta _q } } } \\
 \,\,\,\,\,\,\,+\sum\limits_{n=1}^\infty {\sum\limits_{m=1}^\infty
{\sum\limits_{q=1}^\infty {a_n a_m a_q \omega _m \omega _q k_q ^2\cos \zeta
_n \cos \zeta _m \cos \zeta _q } } } \\
 \,\,\,\,\,\,\,+\sum\limits_{n=1}^\infty {\sum\limits_{m=1}^\infty
{\sum\limits_{q=1}^\infty {a_n a_m a_q \omega _m \omega _q k_m k_q \cos
\zeta _n \sin \zeta _m \sin \zeta _q } } } \\
 \,\,\,\,\,\,\,+\sum\limits_{n=1}^\infty {\sum\limits_{m=1}^\infty
{\sum\limits_{q=1}^\infty {a_n a_m a_q k_q \omega _m \omega _q k_q ^2\cos
\zeta _n \cos \zeta _m \sin \zeta _q } } } \, \\
 \,\,\,\,\,\,\,+\left\{ {\tilde {\phi }_t +\sum\limits_{n=1}^\infty
{\sum\limits_{m=1}^\infty {a_n a_m k_m \omega _m ^2k_m \cos \zeta _n \cos
\zeta _m } } +\sum\limits_{n=1}^\infty {a_n \omega _n \cos \zeta _n \tilde
{\phi }_x } } \right. \\
 \,\,\,\,\,\,\, +\sum\limits_{n=1}^\infty {a_n \omega _n sn\zeta _n \tilde
{\phi }_z } +\sum\limits_{n=1}^\infty {a_n \cos \zeta _n \left( {\tilde
{\phi }_{tz} +O\left( {a_n } \right)} \right)} \\
 \,\,\,\,\,\,   \,+\left. {\frac{1}{2}\sum\limits_{n=1}^\infty
{\sum\limits_{m=1}^\infty {a_n a_m \cos \zeta _n \cos \zeta _m } \left(
{\tilde {\phi }_{tzz} +O\left( {a_n } \right)} \right)} } \right\} _{z=0} =0
\\
 \end{array}
\end{equation}
Neglecting cross terms in (\ref{eq7})
 and (\ref{eq8}) we obtain
for the kinematic condition,
\begin{equation}
\label{eq9}
\begin{array}{l}
 \tilde {\eta }_t +\tilde {\eta }_x \sum\limits_{n=1}^\infty {\omega _n a_n
\cos \zeta _n } -\tilde {\eta }\sum\limits_{n=1}^\infty {\omega _n k_n a_n
\sin \zeta _n } \\
\,\,\,\,\,\, \, -\left( {\tilde {\phi }_z +\tilde {\phi }_x \sum\limits_{n=1}^\infty
{a_n k_n \sin \zeta _n  } +\tilde {\phi }_{zz} \sum\limits_{n=1}^\infty
{a_n \cos \zeta _n } } \right)_{z=0} =0 \\
 \,\, \\
 \end{array}
\end{equation}
and for the dynamic condition,
\begin{equation}
\label{eq10}
\begin{array}{l}
 g\tilde {\eta }-\tilde {\eta }\sum\limits_{n=1}^\infty {\omega _n ^2a_n
\cos \zeta _n } \\
\,\,\,\,\,\,    \,+\left\{ {\tilde {\phi }_t +\tilde {\phi }_x \sum\limits_{n=1}^\infty {a_n
\omega _n \cos \zeta _n } \,+\tilde {\phi }_z \sum\limits_{n=1}^\infty {a_n
\omega _n \sin \zeta _n } +\tilde {\phi }_{tz} \sum\limits_{n=1}^\infty {a_n
\cos \zeta _n } } \right\}_{z=0} =0 \\
 \end{array}
\end{equation}
Equations (\ref{eq9}) and (\ref{eq10}) generalize the boundary
conditions obtained by Benjamin and Feir [1] insofar as terms
corresponding to $n=1$ produce the $O\left( a \right)$ terms they
obtained by a similar perturbation analysis. In that work, the
$O\left( a \right)$ terms were used to specify the form of the
disturbance, while $O\left( {a^2} \right)$ terms were used to
produce coupled equations for the amplitude of the sideband modes.
Neglecting cross terms in (\ref{eq7}) and (\ref{eq8}) eliminated
terms which would correspond to these $O\left( {a^2} \right)$
terms in Benjamin and Feir's work.

\section{Conditions for instability due to a particular harmonic}

While equations (\ref{eq9}) and (\ref{eq10}) were obtained by
ignoring certain cross terms, the form of the equations
generalized the $O\left( a \right)$ terms obtained by Benjamin and
Feir [1]\textbf{. }We now retain precisely the collection of
cross-terms in (\ref{eq7}) and (\ref{eq8}) to produce $O\left(
{a_n ^2} \right)$ terms that will generalize the $O\left( {a^2}
\right)$ in that work. In this way, the kinematic condition may be
represented as
\begin{equation}
\label{eq11}
\begin{array}{l}
 \tilde {\eta }_t -\left( {\tilde {\phi }} \right)_{z=0}
=\sum\limits_{n=1}^\infty {a_n \left\{ {k_n \omega _n \sin \zeta _n \,\tilde
{\eta }-\omega _n \cos \zeta _n \,\tilde {\eta }_x +\left( {k_n \sin \zeta
_n \tilde {\phi }_x +\cos \zeta _n \tilde {\phi }_{zz} } \right)_{z=0} }
\right\}} \\
\,\,\,\,\,\,    \,  +\frac{1}{2}\sum\limits_{n=1}^\infty {a_n^2 \left( {2k_n^2 \omega _n \sin
2\zeta _n \,\tilde {\eta }-k_n \omega _n \left( {1+2\cos 2\zeta _n }
\right)\tilde {\eta }_x } \right)} \\
\,\,\,\,\,\,    \,+\frac{1}{2}\sum\limits_{n=1}^\infty {a_n^2 } \left\{ {k_n \sin 2\zeta
_n \left( {2k_n \tilde {\phi }_x +\tilde {\phi }_{xz} } \right)+k_n \cos
2\zeta _n \tilde {\phi }_{zz} +\frac{1}{2}\left( {1+\cos 2\zeta _n }
\right)\tilde {\phi }_{zzz} } \right\}_{z=0} \\
 \end{array}
\end{equation}
and the dynamic condition represented as
\begin{equation}
\label{eq12}
\begin{array}{l}
 g\tilde {\eta }+\left( {\tilde {\phi }_t } \right)_{z=0}
=\sum\limits_{n=1}^\infty {a_n \left( {\omega _n ^2\cos \zeta _n \,\tilde
{\eta }-\left( {\omega _n \cos \zeta _n \,\tilde {\phi }_x +\omega _n \sin
\zeta _n \,\tilde {\phi }_z +\cos \zeta _n \,\tilde {\phi }_{tz} }
\right)_{z=0} } \right)}  \\
\,\,\,\,\,\,    \,  +\frac{1}{2}\sum\limits_{n=1}^\infty {a_n^2 \left( {k_n \omega _n^2 \left(
{1-\cos 2\zeta _n } \right)\,\tilde {\eta }+\left( {\omega _n \sin 2\zeta _n
\left( {k_n \tilde {\phi }_z +\tilde {\phi }_{zz} } \right)_{z=0} } \right)}
\right)} \\
\,\,\,\,\,\,    \,  +\frac{1}{2}\sum\limits_{n=1}^\infty {a_n^2 \left( {\left( {1+\cos 2\zeta
_n } \right)\left( {k_n \omega _n \tilde {\phi }_x +\omega _n \tilde {\phi
}_{xz} +\frac{1}{2}\tilde {\phi }_{zzt} } \right)+k_n \cos 2\zeta _n
\,\tilde {\phi }_{zt} } \right)_{z=0} } \\
 \end{array}
\end{equation}
Now, assume that the disturbance consists of two sideband modes together
with the products of their interaction with the basic wave train,
\[
\tilde {\eta }=\tilde {\eta }_1 +\tilde {\eta }_2 \quad ,
\quad
\tilde {\phi }=\tilde {\phi }_1 +\tilde {\phi }_2 .
\]
We define the arguments
\begin{equation}
\label{eq13} \zeta _{n,1} =k_n \left( {1+\kappa } \right)x-\omega
_n \left( {1+\delta } \right)t-\gamma _{n,1} \left( t \right)
\end{equation}
\begin{equation}
\label{eq14} \zeta _{n,2} =k_n \left( {1-\kappa } \right)x-\omega
_n \left( {1-\delta } \right)t-\gamma _{n,2} \left( t \right),
\end{equation}
where $\gamma _{n,1} \left( t \right)$ and $\gamma _{n,2} \left( t
\right)$ are required so that each harmonic satisfies the
dispersion relation to $O\left( {\delta ^2} \right)$. We denote
frequencies and wave numbers for the sidebands by \be \label{eq15}
& &\omega _{n,1} =\omega \left( {1+\delta } \right) \label{eq15}
\\
& &\omega _{n,2} =\omega \left( {1-\delta } \right),\quad
\en
and
\begin{equation}
\label{eq16}
k_{n,1} =k_n \left( {1+\kappa } \right)
\end{equation}
\begin{equation}
\label{eq17}
k_{n,2} =k_n \left( {1-\kappa } \right)
\end{equation}
where $\kappa $ and $\delta $ are small constants.

Then, for $i=1,2$, we define
\begin{equation}
\label{eq18}
\tilde {\eta }_i =\sum\limits_{n=1}^\infty {\varepsilon _{n,i} \cos \zeta
_{n,i} } +\sum\limits_{n=1}^\infty {\varepsilon _{n,i} k_n a_n \cos
\left( {\zeta _n +\zeta _{n,i} } \right)+O\left( {k_n^2 a_n^2 \varepsilon
_{n,i} } \right)}
\end{equation}
and
\be
 \tilde {\phi }_i &=&\sum\limits_{n=1}^\infty {k_{n,i} ^{-1}e^{k_{n,i}
z}\left\{ {\varepsilon _{n,i} \left( {\omega _{n,i} L_{n,i} +\dot {\gamma
}_{n,i} M_{n,i} } \right)\sin \zeta _{n,i} +\dot {\varepsilon }_{n,i}
N_{n,i} \cos \zeta _{n,i} } \right\}}\no \\
    &+&\sum\limits_{n=1}^\infty {\omega _n a_n \varepsilon _{n,i} D_{n,i}
e^{\left| {k_n -k_{n,i} } \right|z}\sin \left( {\zeta _n -\zeta _{n,i} }
\right)}\label{eq19}
\en
We further assume that $\varepsilon _{n,i} $and $\gamma _{n,i} $ are slowly
varying functions of time, such that their derivatives have the properties
\begin{equation}
\label{eq20}
\dot {\varepsilon }_{n,i} =O\left( {w_n k_n ^2a_n ^2\varepsilon_{n,i} }
\right),\quad
\quad
\dot {\gamma }_{n,i} =O\left( {w_n k_n ^2a_n ^2} \right).
\end{equation}
Now, among the infinitely many products arising from the nonlinear
interaction between these disturbance modes and the basic wave train, there
will be components with arguments
\begin{equation}
\label{eq21}
\left.\begin{array}{l}
 2\zeta _n -\zeta _{n,1} =\zeta _{n,2} +\left( {\gamma _{n,1} +\gamma _{n,2}
} \right) \\

 2\zeta _n -\zeta _{n,2} =\zeta _{n,1} +\left( {\gamma _{n,1} +\gamma _{n,2}
} \right) \\
 \end{array}\right\}
\end{equation}
suggesting that resonance may be induced between the sidebands by the
interaction of particular harmonics. This may contribute to instability if
the sum of the time-dependent phase functions approaches a constant,
\be
\theta _n =\gamma _{n,1} +\gamma _{n,2} \to \mbox{const. as}\quad t\to \infty\label{21}
\en

Indeed, upon substitution of (\ref{eq18}) and (\ref{eq19}) into
(\ref{eq9}) and (\ref{eq10}), a variety of cross terms involving
interaction between the primary wave and the disturbance may be
identified so that (\ref{eq21}) applies. For example,
\[
\sin 2\zeta _n \cos \zeta _{n,1} =\tf{1}{2}\left[ {\sin \left( {2\zeta _n
-\zeta _{n,1} } \right)+\sin \left( {2\zeta _n +\zeta _{n,1} } \right)}
\right]\sim \tf{1}{2}\sin \left( {\zeta _{n,2} +\theta _n } \right)
\]
and similarly
\[
\sin 2\zeta _n \cos \zeta _{n,2} =\tf{1}{2}\left[ {\sin \left( {2\zeta _n
-\zeta _{n,2} } \right)+\sin \left( {2\zeta _n +\zeta _{n,2} } \right)}
\right]\sim \tf{1}{2}\sin \left( {\zeta _{n,1} +\theta _n } \right).
\]
Upon substitution, there exists a set of interactions so that, after
equating coefficients, the kinematic boundary condition reduces to the pair
of equations
\begin{equation}
\label{eq22}
\begin{array}{l}
 \varepsilon _{n,1} \left\{ {\omega _{n,1} \left( {1-L_{n,1} } \right)+\dot
{\gamma }_{n,1} \left( {1-M_{n,1} } \right)} \right\}\sin \zeta _{n,1} +\dot
{\varepsilon }_{n,1} \left( {1-N_{n,1} } \right)\cos \zeta _{n,1} \\
 \,\,\,\,\,\, \,    =\omega _n k_n^2 a_n^2 \left\{ {\frac{5}{4}\varepsilon _{n,1} \sin \zeta
_{n,1} +\frac{5}{8}\varepsilon _{n,2} \sin \left( {\zeta _{n,1} +\theta _n }
\right)} \right\} \\
 \end{array}
\end{equation}
\begin{equation}
\label{eq23}
\begin{array}{l}
 \varepsilon _{n,2} \left\{ {\omega _{n,2} \left( {1-L_{n,2} } \right)+\dot
{\gamma }_{n,2} \left( {1-M_{n,2} } \right)} \right\}\sin \zeta _{n,2} +\dot
{\varepsilon }_{n,2} \left( {1-N_{n,2} } \right)\cos \zeta _{n,2} \\
\,\,\,\,\,\,  \,    =\omega _n k_n^2 a_n^2 \left\{ {\frac{5}{4}\varepsilon _{n,2} \sin \zeta
_{n,2} +\frac{5}{8}\varepsilon _{n,1} \sin \left( {\zeta _{n,2} +\theta _n }
\right)} \right\} \\
 \end{array}.
\end{equation}
for $n=1,2,3,...$. Furthermore, there exists a set of interactions so that
the dynamic boundary condition reduces to the pair of equations
\begin{equation}
\label{eq24}
\begin{array}{l}
 \varepsilon _{n,1} \left\{ {\omega _{n,1}^{-1} \left( {gk_{n,1} -\omega
_{n,1}^2 L_{n,1} } \right)-\dot {\gamma }_{n,1} \left( {1+M_{n,1} } \right)}
\right\}\cos \zeta _{n,1} +\dot {\varepsilon }_{n,1} \left( {1+N_{n,1} }
\right)\sin \zeta _{n,1} \\
\,\,\,\,\,\,  \,    =-\omega k_n^2 a_n^2 \left\{ {\frac{3}{4}\varepsilon _{n,1} \cos \zeta
_{n,1} +\frac{3}{8}\varepsilon _{n,2} \cos \left( {\zeta _{n,1} +\theta _n }
\right)} \right\} \\
 \end{array}
\end{equation}
and
\begin{equation}
\label{eq25}
\begin{array}{l}
 \varepsilon _{n,2} \left\{ {\omega _{n,2}^{-1} \left( {gk_{n,2} -\omega
_{n,2}^2 L_{n,2} } \right)-\dot {\gamma }_{n,2} \left( {1+M_{n,2} } \right)}
\right\}\cos \zeta _{n,2} +\dot {\varepsilon }_{n,2} \left( {1+N_{n,2} }
\right)\sin \zeta _{n,2} \\
\,\,\,\,\,\,  \,    =-\omega k_n^2 a_n^2 \left\{ {\frac{3}{4}\varepsilon _{n,2} \cos \zeta
_{n,2} +\frac{3}{8}\varepsilon _{n,1} \cos \left( {\zeta _{n,2} +\theta _n }
\right)} \right\} \\
 \end{array}
\end{equation}
for $n=1,2,3,...$. By adding the coefficients of $\cos \zeta _{n,i} $ in (22)
and coefficients of $\sin \zeta _{n,i} $in (23); then, adding coefficients
of $\sin \zeta _{n,i} $ in (22) and $\cos \zeta _{n,i} $ in (23), we obtain
a system of four equations,
\begin{equation}
\label{eq26}
\left. {\begin{array}{l}
 \df{d\varepsilon _{n,1}
}{dt}=\left( {\frac{1}{2}\omega k_n ^2a_n ^2\sin \theta _n }
\right)\varepsilon _{n,2} \\
 \\
 \df{d\varepsilon _{n,2} }{dt}=\left(
{\frac{1}{2}\omega k_n ^2a_n ^2\sin \theta _n } \right)\varepsilon _{n,1} \\
\\
 \df{d\gamma _{n,1} }{dt}=\frac{1}{2}\left( {\frac{gk_{n,1} }{\omega
_{n,1} }-\omega _{n,1} } \right)+\omega _{n,1} k_{n,1} ^2a_n ^2\left(
{1+\frac{\varepsilon _{n,2} }{2\varepsilon _{n,1} }\cos \theta _n } \right) \\
\\
 \df{d\gamma _{n,2} }{dt}=\frac{1}{2}\left( {\frac{gk_{n,2} }{\omega
_{n,2} }-\omega _{n,2} } \right)+\omega _{n,2} k_{n,2} ^2a_n ^2\left(
{1+\frac{\varepsilon _{n,1} }{2\varepsilon _{n,2} }\cos \theta _n } \right)
\\
 \end{array}} \right\}
\end{equation}
The last two equations of (\ref{eq26}) may be added to give an equation for $\theta
_n =\gamma _{n,1} +\gamma _{n,2} $,
\begin{equation}
\label{eq27}
\frac{d\theta _n }{dt}=\omega _n k_n ^2a_n ^2\left\{ {1+\frac{\varepsilon
_{n,1} ^2+\varepsilon _{n,2} ^2}{2\varepsilon _{n,1} \varepsilon _{n,2}
}\cos \theta _n } \right\}-\omega _n \delta ^2
\end{equation}
The resulting three equations in $\varepsilon _{n,1} $, $\varepsilon _{n,2}
$, and $\theta _n $ may be reduced to a single equation in $\varepsilon
_{n,1} $ by introducing the parameters
\begin{equation}
\label{eq28}
T_n =k_n ^2a_n ^2\omega t
\end{equation}
and
\begin{equation}
\label{eq29}
\alpha _n =1-\frac{\delta ^2}{k_n^2 a_n^2 }.
\end{equation}
It may then be shown [1],
\begin{equation}
\label{eq30}
\frac{d\varepsilon _{n,1}^{^2} }{dT_n }=\frac{d\varepsilon _{n,2}^2 }{dT_n
}=\varepsilon _{n,1} \varepsilon _{n,2} \sin \theta _n \quad ,
\end{equation}
\begin{equation}
\label{eq31}
\varepsilon _{n,1} \varepsilon _{n,2} \cos \theta _n +\alpha _n \varepsilon
_{n,1}^2 =\rho _n =const,
\end{equation}
and
\begin{equation}
\label{eq32}
\varepsilon _{n,1}^2 -\varepsilon _{n,2}^2 =2\alpha _n \rho _n \left( {1-v_n
} \right)=const.
\end{equation}
We then obtain
\begin{equation}
\label{eq33}
\left( {\frac{d\varepsilon _{n,1}^2 }{dT_n }} \right)^2=\left( {1-\alpha _n
^2} \right)\varepsilon _{_{n,1} }^4 +2\alpha _n v_n \rho _n \varepsilon
_{_{n,1} }^2 -\rho _n ^2.
\end{equation}
Equation (\ref{eq33}) shows that the growth rate of the sidebands
is expressible as a quadratic in $\varepsilon _{n,1}^2 $. We first
consider the case $\alpha _n^2 =1$, when (\ref{eq33}) reduces to a
linear function in $\varepsilon _{n,1}^2 $. Since (\ref{eq29})
implies $\alpha _n <1$, we consider $\alpha _n =-1$. Since
(\ref{eq33}) must be satisfied for arbitrary initial values of
${d\varepsilon _{n,1}}/{dT_n }$ and $\varepsilon _{n,1}^2 $, it
can be shown that $\rho _n v_n <0$ so that (\ref{eq33}) may be
solved directly, with $\varepsilon _{n,1}^2 \sim T_n $ so that
this case must correspond to instability\textbf{. } We now suppose
$\alpha _n^2 \ne 1$ and consider four cases for the form of the
quadratic function (\ref{eq33}), as shown in Figure 1.

\begin{figure}
   \begin{center}
\includegraphics[width=11.5cm]{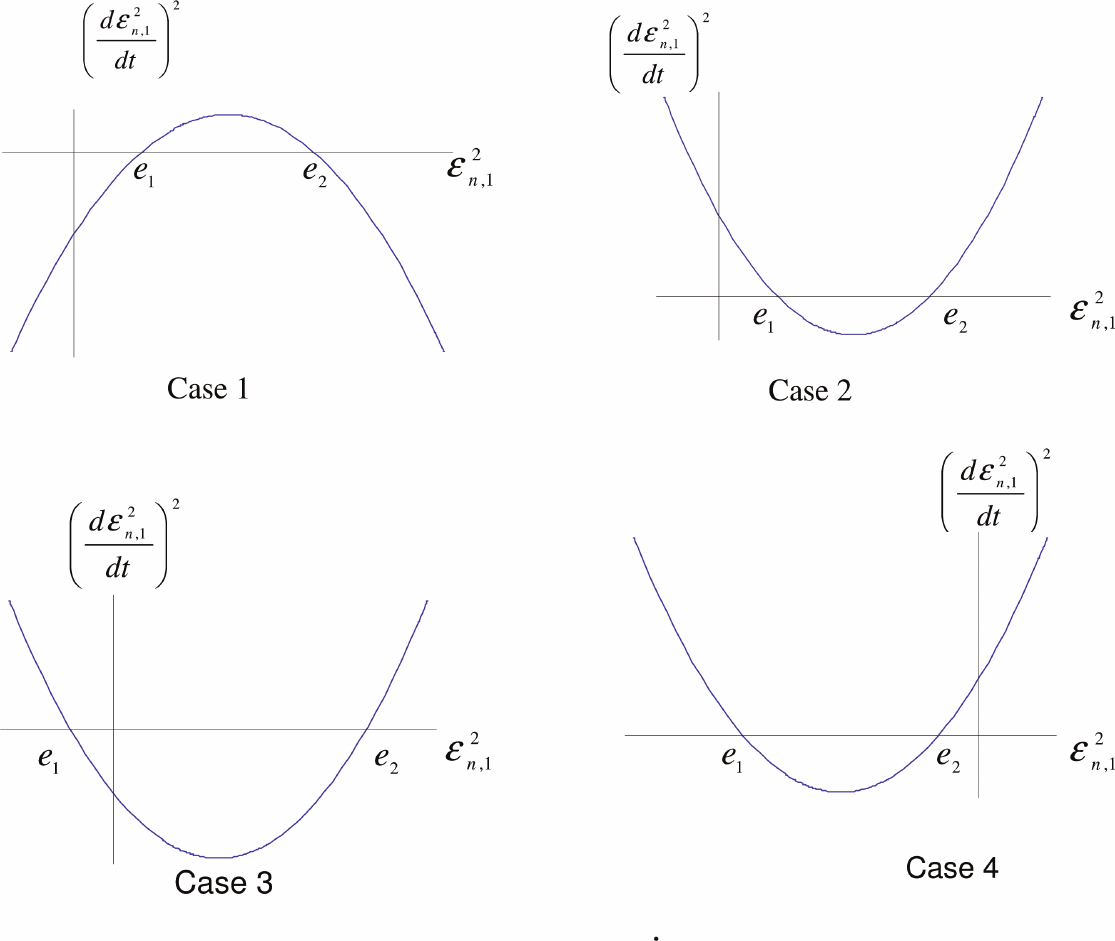}
   \end{center}
\caption{\footnotesize Growth rate of sidebands correspo$e_1 e_2 \left(
{\frac{d\varepsilon _{n,1}^2 }{dt}} \right)^2$nding to $n^{th}$ harmonic.}
\end{figure}

In the case $1-\alpha _{_n }^2 <0$, unbounded sideband growth
cannot occur, as there exists only a finite interval $e_1 \le
\varepsilon _{n,1}^2 \le e_2 $ where the growth rate $\left(
{{d\varepsilon _{n,1}^2 }/{dT_n }} \right)^2>0$. If the initial
conditions at time $T_n =T_{n,0} $ place $\varepsilon _{n,1}^2
\left( {T_{n,0} } \right)$ within this interval, then the
amplitude $\varepsilon _{n,1}^2 \left( {T_n } \right)$ must be
greater at a slightly greater time $T$. Then $\varepsilon _{n,1}^2
\left( {T_n } \right)$ will increase until it reaches the value
$e_2 $, when $\left( {{d\varepsilon _{n,1}^2 }/{dT_n }}
\right)^2=0$ and further growth is suppressed. If the initial
conditions at $T_n =T_{n,0} $ place $\varepsilon _{n,1}^2 \left(
{T_{n,0} } \right)$ outside this interval, then the amplitude
$\varepsilon _{n,1}^2 \left( {T_n } \right)$ will decrease at
slightly greater time $T$, and unbounded growth will not occur.

We therefore consider the case $1-\alpha _{_n }^2 \ge 0$ which, from (\ref{eq29})
corresponds to
\begin{equation}
\label{eq34}
0\le \delta <\sqrt 2 k_n a_n ,
\end{equation}
providing the frequency range for the classical Benjamin-Feir instability in
the case $n=1$, when $k_1 =k$ and $a_1 =a$. If a higher harmonic satisfies
(\ref{eq34}), then (\ref{eq33}) shows there exists an infinite interval $e_1 <\varepsilon
_{n,1}^2 $ where the growth rate $\left( {{d\varepsilon _{n,1}^2 }/{dT_n
}} \right)^2>0$. For each harmonic, if $1-\alpha _n ^2>0$, three cases arise
depending on the number of intercepts on the positive $\varepsilon _{n,1}^2
$ axis (cases 2-4 in Figure 1). Arguing as in case 1, the initial
condition $\varepsilon _{n,1}^2 \left( {T_{n,0} } \right)>e_2 $ insures
unbounded growth in cases 2 and 3, while any initial condition leads to
unbounded growth in case 4. With these conditions, the wave is unstable
regardless of the sign of $\varepsilon _{n,1} \left( {T_{n,0} } \right)$,
and the asymptotic growth rate of both sidebands is equal from (\ref{eq32}).

We can characterize the conditions leading to unbounded growth due to the
interaction between the primary wave and the $n^{th}$ harmonic of the
disturbance by noting that (\ref{eq33}) may be written in the form
\begin{equation}
\label{eq35}
\left( {\frac{d\varepsilon _{n,1}^2 }{dT_n }} \right)^2=\left( {1-\alpha _n
^2} \right)\left\{ {\left( {\varepsilon _{_{n,1} }^2 -A_N } \right)^2-B_n
^2} \right\},
\end{equation}
where
$$A_n =-\frac{\alpha _n v_n \rho _n }{1-\alpha _n^2 }\quad {\rm and}\quad B_n
=\frac{\rho _n \left( {1-\alpha _n^2 +\alpha _n^2 v_n^2 } \right)^{1
\mathord{\left/ {\vphantom {1 2}} \right. \kern-\nulldelimiterspace}
2}}{\vert 1-\alpha _n^2 \vert },$$
so that in Figure 1
\begin{equation}
\label{eq36}
e_2 =\max \left\{ {A_n +B_n ,A_n -B_n } \right\},
\end{equation}
and the condition for unbounded growth in cases 2,3, and 4 may be written as
\begin{equation}
\label{eq37} \varepsilon _{n,1} \left( {T_{n,0} } \right)>e_2
\end{equation}
If (\ref{eq34}) and (\ref{eq37}) both hold, then unbounded growth of the $n^{th}$ harmonic
of the disturbance will occur. Furthermore, the sign of $\varepsilon _{n,1}
\left( {T_n } \right)$ will not change from its sign at initial time $T_0 $,
and inspection of (\ref{eq26}) shows that $\varepsilon _{n,1} \left( {T_n } \right)$
and $\varepsilon _{m,1} \left( {T_{m,0} } \right)$ must have the same sign.
Therefore, it is not possible for unbounded sideband growth to cancel, when
a pair of sidebands for a single harmonic is considered.

\section{Conditions for instability: Collective effects of all harmonics}

In Section 3, conditions were given on the disturbance and the
initial conditions so that interaction between the primary wave
and the $n^{th}$ harmonic of the disturbance produce unbounded
growth of the sideband amplitudes $\varepsilon _{n,1} \left( {T_n
} \right)$ and $\varepsilon _{n,2} \left( {T_n } \right)$. These
conditions were dependent on the coefficients of the Stokes wave,
$\eta \left( {x,t} \right)=\sum_{n=1}^\infty {a_n \cos \zeta _n }
$. Longuet-Higgins showed that the Stokes coefficients are
decreasing up to very high orders of $n$ [2,3]. In that
representation, each coefficient is represented by a series where
the leading term agrees with a formula of Wilton [5],
\begin{equation}
\label{eq38}
a_n =\frac{n^n}{n!}s_n ,
\end{equation}
with $s_n $ a parameter related to the wave height. The first few terms of
the Stokes wave may then be written
\[
y=s_1 \cos kx+2s_2 \cos 2kx+\tf{27}{6}s_3 \cos 3kx+\tf{64}{24}s_4 \cos
4kx+\tf{3125}{120}s_5 \cos 5x+...
\]
Using Stokes' representation [4],
\[
y=a\cos kx+\tf{1}{2}ak^2\cos 2kx+\tf{3}{8}a^3k^2\cos
3kx+\tf{1}{3}a^4k^3\cos 4kx+\tf{125}{384}a^5k^4\cos 5x+....
\]
and comparing coefficients, we have
$$s_1 =a,\quad s_2 =\tf{1}{4}a^2k,\quad s_3
=\tf{1}{12}a^3k^2, \quad s_4 =\tf{1}{8}a^3k^2,\quad s_5
=\tf{24}{625}a^3k^2.$$

Now consider the collective contribution of all harmonics in
(\ref{eq18}) and (\ref{eq19}). As noted in Section 3, values of
$\delta \left( n \right)$ providing the marginal stability case
for a particular harmonic decrease with $n.$ Therefore, if a
particular harmonic of the disturbance has frequency sufficiently
close to the corresponding harmonic of the primary wave, unbounded
sideband growth will occur for all lower order harmonics, so long
as the initial condition (\ref{eq37}) is met. Furthermore, if
(\ref{eq37}) holds for $n=M$, but not for $n=M+1$, then unbounded
growth will indeed occur, as the coefficients of the higher
harmonics will decay in time and will not suppress the
instability.

\begin{figure}
   \begin{center}
\includegraphics[width=11.5cm]{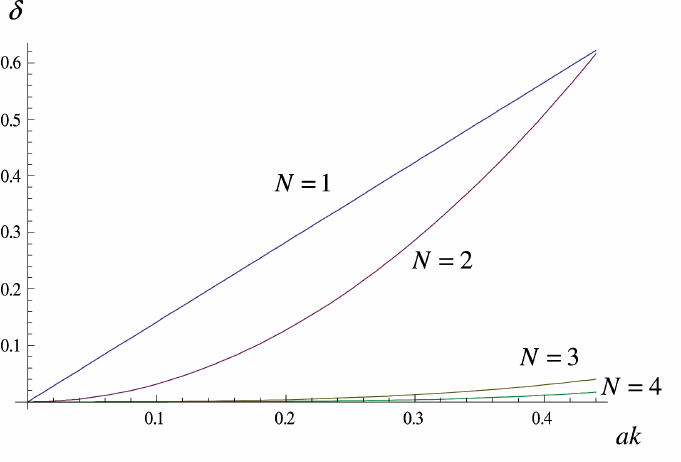}
   \end{center}
\caption{\footnotesize The maximum value of $\delta $ as a function of wave
steepness \textit{ak} so that all harmonics up to the $N^{th}$ harmonic contribute to
instability.
$ak
\delta
N=2
N=1
N=3
N=4$.}
\end{figure}

Now suppose that (\ref{eq34}) holds for some harmonic. Then it
must hold for all lower order harmonics, and we may consider the
sequence of harmonics where (\ref{eq34}) holds. For these
harmonics, since only cases 2-4 in Figure 1 need be considered,
\[
-1<\alpha _n <1
\]
or, equivalently,
\[
0<\frac{\delta ^2}{n^2k^2a_n^2 }<2 \] From (\ref{eq38}),
\[
0<\frac{\left( {n!} \right)^2\delta ^2}{n^2k^2n^{2n}s_n^2 }<2
\]
which may be written
\begin{equation}
\label{eq39}
0<\delta <\frac{\sqrt 2 \,k\,n^ns_n }{\left( {n-1} \right)!}.
\end{equation}
For the first few harmonics, the marginal stability condition is,
\be
\begin{array}{lcl}
n=1: & &
0<\delta <\df{\sqrt 2 \,k\,n^ns_n }{\left( {n-1} \right)!}=\sqrt 2 ak\\
\\
n=2: & &
0<\delta <\df{\sqrt 2 \,k\,n^ns_n }{\left( {n-1} \right)!}=4\sqrt 2
\,k\,s_2 =\sqrt 2 \,a^2k^2\\
\\
n=3: & &
0<\delta <\df{\sqrt 2 \,k\,n^ns_n }{\left( {n-1} \right)!}=\df{27\sqrt 2
\,ks_3 }{4}=\df{9\sqrt 2 }{16}\,a^3k^3\\
\\
n=4: & &
0<\delta <\df{\sqrt 2 \,k\,n^ns_n }{\left( {n-1} \right)!}=\df{256\sqrt
2 \,ks_4 }{6}=\df{16\sqrt 2 }{3}\,a^4k^4\\
\\
n=5: & &
0<\delta <\df{\sqrt 2 \,k\,n^ns_n }{\left( {n-1} \right)!}=\df{3125\sqrt
2 \,ks_4 }{24}=5\sqrt 2 \,a^5k^5\\
\\
\end{array}\no
\en
Since $ak\ll 1$, the value of $\delta $ corresponding to the marginal
instability decreases with $n$. This is expected, since disturbance frequencies
sufficiently close to that of the primary wave so that sideband amplitude
$\varepsilon _{n,1} $ becomes unbounded, will be sufficiently small so that
sideband amplitudes for all lower order harmonics are also unbounded.
Therefore, (\ref{eq39}) may be considered a generalization of the Benjamin-Feir
criterion in the sense that, if it is satisfied, all harmonics up to order
$n$ contribute to unbounded growth. This conclusion is subject to initial
conditions, represented by (\ref{eq37}). Indeed, if (\ref{eq39}) is satisfied for a
(possibly infinite) set of harmonics, instability will occur only if (\ref{eq37}) is
satisfied for at least one value of $n$.

Values for the first five coefficients using wavelength ${2\pi }
\mathord{\left/ {\vphantom {{2\pi } k}} \right.
\kern-\nulldelimiterspace} k=64$ cm and $a=0.1$ are compiled in
Table 1 using Wilton's formula for the coefficients. These
parameters give $ak=0.098\ll 1$. The last column provides the
marginal instability criterion given in (\ref{eq39}).

\begin{table}
  \begin{center}
  \begin{tabular}{ccccccc}
$n$ & $s_n$          & $a_n =\left( {\frac{n^n}{n!}} \right)s_n$   & $\delta =\frac{\sqrt 2 \,k\,n^ns_n }{\left( {n-1} \right)\,!}$ \\[3pt]
1   & $a$            & 0.1                                         & 0.013884009\\
2   & $a^2k/4$       & 0.000490874                                 & 0.000136306\\
3   & $a^3k^2/12$    & 0.003681554                                 & 0.001533442\\
4   & $a^4k^3/8$     & $1.26165\times 10^{-7}$                     & $7.0067\times 10^{-8}$\\
5   & $24a^5k^4/625$ & $9.28965\times 10^{-10}$                    & $6.4488\times 10^{-10}$\\
  \end{tabular}
  \caption{Sample coefficients of Stokes wave and disturbance
frequency corresponding to marginal instability for the n$^{th}$ harmonic.}
  \label{tab:kd}
  \end{center}
\end{table}

Equation (\ref{eq38}) shows that the instability criterion
(\ref{eq39}) depends on the wave steepness $a^nk^{n-1}$. In Figure
2, we illustrate values of $\delta \left( n \right)$ as a function
of steepness for $1\le n\le 4$ Note that the value of $\delta
\left( n \right)$ decreases with $n$ for all values of $ak<.44$.
While $\delta \left( 1 \right)=\delta \left( 2 \right)$ for
$ak=0.44$, it must be remembered that the small amplitude
assumption $ak<<1$ is in place, so that values of \textit{ak}
where this equality occurs are arguably outside the realm
applicable to this analysis.

\begin{figure}
   \begin{center}
\includegraphics[width=11.5cm]{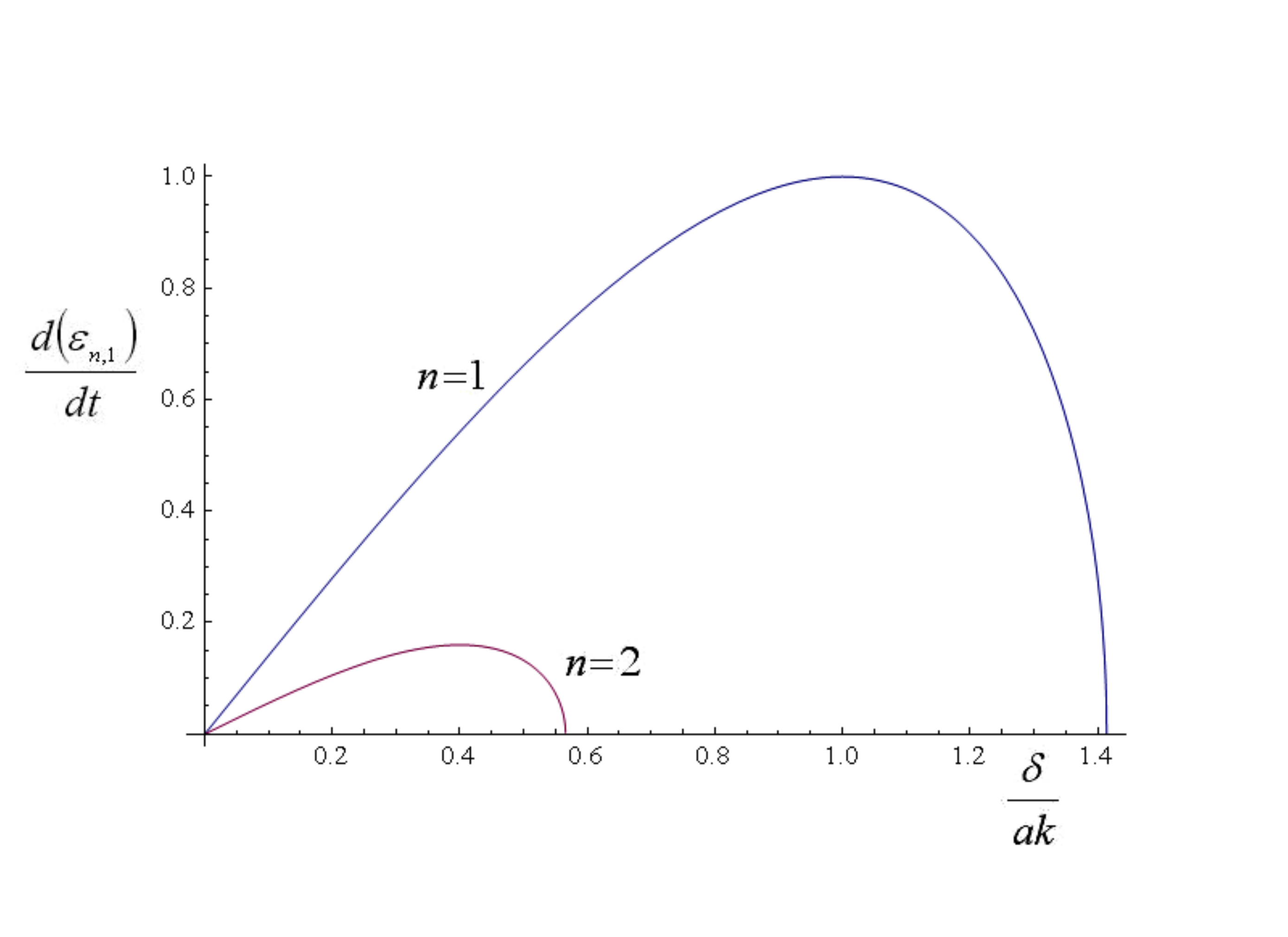}
   \end{center}
\caption{\footnotesize Growth rate ${d}\ln \left( {\varepsilon _{n,1} }
\right)/{dt}$ as a function of ${\delta }/{ak}$ for cases $n=1,2,3$.}
\end{figure}

We now consider the growth of the amplitude of the disturbance.
Following [1], we solve (\ref{eq35}) directly for ${d\varepsilon
_{n,1}^2 }/{dT_n }$ by separating variables, and recalling the
asymptotic equivalence of $\varepsilon _{n,1} \left( {T_n }
\right)$ and $\varepsilon _{n,2} \left( {T_n } \right)$. We obtain
\begin{equation}
\label{eq40}
\varepsilon _{n,1} \sim \exp \left[ {\frac{\delta }{2}\sqrt {2k_n^2 a_n^2
-\delta ^2} \,\,\omega T_n } \right].
\end{equation}
Applying (\ref{eq28}),
\begin{equation}
\label{eq41}
\ln \varepsilon _{n,1} =\frac{\delta }{2}\omega \,a_n k_n \sqrt {2-\left(
{\frac{\delta }{a_n k_n }} \right)^2} \,\,\,t,
\end{equation}
and then (\ref{eq38}), with $k_n =nk$,
\begin{equation}
\label{eq42}
\ln \varepsilon _{n,1} =\frac{\delta }{2}\omega \,\left( {\frac{n^n}{n!}s_n
} \right)nk\sqrt {2-\left( {\frac{n!\delta }{n^{n+1}ks_n }} \right)^2}
\,\,\,t.
\end{equation}
If we select $s_n =a^nk^{n-1}$, noting that this parameter goes to
zero with the wave height as required, we obtain
\begin{equation}
\label{eq43}
\ln \varepsilon _{n,1} =\frac{\omega a^2k^2}{2}\,\left( {\frac{n^{n+1}}{n!}}
\right)\left( {\frac{\delta }{ak}} \right)\sqrt {2-\left(
{\frac{(n-1!}{n^n(ak)^{n-1}}} \right)^2\left( {\frac{\delta }{ak}}
\right)^2} \,\,\,t.
\end{equation}
Figure 3 shows a graph of ${d}\ln \left( {\varepsilon _{n,1} }
\right)/{dt}$ versus ${\delta}/{ak}$ where the case $n=1$
corresponds to Figure 1 in [1] and, following that development, we
have employed the normalization ${wa^2k^2}/{2}=1$. For $n>1$, the
factor \textit{ak }appears and Figure 1 uses $ak=0.1\ll 1$ to
display the growth rate of the amplitude for the first two
harmonics. The pattern persists for higher harmonics, and
(\ref{eq43}) gives the maximum growth rate occurring at
\begin{equation}
\label{eq44}
\delta =\frac{n\left( {ak^{n-1}} \right)}{\left( {n-1} \right)!}.
\end{equation}
Note that in the case $n=1$, (\ref{eq44}) gives the maximum growth
rate occurring at the known value of $\delta =ak$, while the
maximum growth rate of the second harmonic occurs at $\delta
=2ak$, shown in Figure 3 for the case $ak=0.1.$

Assuming appropriate initial conditions and sufficiently small
$\delta$ so that a set of initial harmonics contribute to
resonance, the collective asymptotic growth rate represented by
the $N^{th}$ partial sum will be approximated by the sum of the
growth rates of the individual harmonics. From (\ref{eq42}), we
have

\begin{equation}
\label{eq45}
\sum\limits_{n=1}^N {\frac{d\varepsilon _{n,1} }{dt}=\sum\limits_{i=1}^N
{\frac{\delta }{2}\omega a_n k_n \sqrt {2n^2-\left( {\frac{\delta }{a_n k_n
}} \right)^2} \,\,\,\exp \left\{ {\frac{\delta }{2}\omega a_n k_n \sqrt
{2n^2-\left( {\frac{\delta }{a_n k_n }} \right)^2} } \right\}t} }.
\end{equation}

which, for sufficiently large $N$, represents the asymptotic
growth rate for the fully nonlinear Stokes wave.

\end{document}